\documentclass [] {aa} 
\usepackage{epsfig, graphicx,longtable}
\usepackage{natbib}
\usepackage{times}

\bibpunct{(}{)}{;}{a}{}{,}

\begin{document}


\title{Abundances of Na, Mg and Al in stars with giant planets\thanks{Based 
	       on observations collected at the La Silla Observatory, 
               ESO (Chile), with the CORALIE spectrograph 
               at the 1.2-m Euler Swiss telescope and the FEROS spectrograph
               at the 1.52-m and 2.2-m ESO telescopes, with the VLT/UT2 
               Kueyen telescope (Paranal Observatory, ESO, Chile) using the 
               UVES spectrograph (Observing run 67.C-0206, in service 
               mode), with the TNG and William Herschel Telescopes, both 
	       operated at the island of La Palma, and with the ELODIE 
	       spectrograph at the 1.93-m telescope at the Observatoire 
	       de Haute Provence.}}


\author{P.~Beir\~ao\inst{1} \and
	N.C.~Santos\inst{1,2} \and 
        G.~Israelian\inst{3} \and 
	M.~Mayor\inst{2}
	}

\offprints{Nuno C. Santos, \email{Nuno.Santos@oal.ul.pt}}

\institute{
        Centro de Astronomia e Astrof{\'\i}sica da Universidade de Lisboa,
        Observat\'orio Astron\'omico de Lisboa, Tapada da Ajuda, 1349-018
        Lisboa, Portugal
     \and
	Observatoire de Gen\`eve, 51 ch.  des 
	Maillettes, CH--1290 Sauverny, Switzerland
     \and
	Instituto de Astrof{\'\i}sica de Canarias, E-38200 
        La Laguna, Tenerife, Spain
        }
	
\date{Received / Accepted } 

\titlerunning{Na, Mg and Al in stars with giant planets} 


\abstract{
We present Na, Mg and Al abundances in a set of 98 stars with known giant
planets, and in a comparison sample of 41 ``single'' stars. The results
show that the [X/H] abundances (with X = Na, Mg and Al) are, on average, higher in 
stars with giant planets, a result similar to the one found for iron.
However, we did not find any strong difference in the [X/Fe] ratios, for
a fixed [Fe/H], between the two samples of stars in the region where
the samples overlap. The data was used to study the Galactic
chemical evolution trends for Na, Mg and Al and to discuss
the possible influence of planets on this evolution. The results,
similar to those obtained by other authors, show that the [X/Fe]
ratios all decrease as a function of metallicity
up to solar values. While for Mg and Al this trend then becomes relatively
constant, for Na we find indications of an upturn up to [Fe/H] values
close to 0.25\,dex. For metallicities above this value the [Na/Fe] becomes 
constant.
}

\maketitle

\section{Introduction}
\label{sec:intro}

The study of the chemical abundances of planet-host stars is providing
important clues about the processes of planet formation and evolution.
Soon after the discovery of the first exoplanets, it had been suggested
that their host stars were particularly metal-rich compared
to average field dwarfs \citep[][]{Gonzalez-1998}. As new planets
were added to the lists, this fact became more and more clear \citep[e.g.][]{Santos-2001,Santos-2003,Santos-2004b,Gonzalez-2001,Reid-2002}.
Today, it is well accepted that the probability of finding a planet is a strongly
increasing function of the stellar metallicity, at least for [Fe/H] values above
solar \citep[][]{Santos-2004a}.

Besides the analysis of the global metal content of the planet-host stars,
some efforts have also been made to analyze 
the chemical abundances of specific elements. 
These include light element abundances 
\citep[e.g.][]{GarciaLopez-1998,Gonzalez-2000,Ryan-2000,Deliyannis-2000,Israelian-2003,
Israelian-2004,Santos-2002c,Santos-2004c}, as well as other metals \citep[e.g.][]{Santos-2000b,
Gonzalez-2000,Gonzalez-2001,Smith-2001,Takeda-2001,
Sadakane-2002,Bodaghee-2003,Ecuvillon-2004a,Ecuvillon-2004b}.
In general, these studies agree that planet host stars are globally metal-rich
compared to single field dwarfs. With a few exceptions \citep[e.g.][]{Gonzalez-2001,Israelian-2003}
no major unexpected trends were suggested.

In their study of Na, Mg and Al abundances in stars harboring giant planets, 
\citet[][]{Gonzalez-2001} have presented evidence
that the ratios [Na/Fe], [Mg/Fe] and [Al/Fe] may be smaller, for a given
[Fe/H], in stars with planets when compared to ``single'' dwarfs. Unfortunately,
\citet[][]{Gonzalez-2001} had to compare their Na, Mg and Al abundances
for planet-hosts with those derived by other authors for field stars. 
The cause of the observed difference, not seen by other authors \citep[][]{Sadakane-2002},
may thus be simply due to the use of non-uniform samples.

To address this issue, we present in this paper a uniform analysis
of the abundances of Na, Mg and Al in a sample of 98 stars with planets
and a comparison sample of 41 ``single'' field stars. Stellar parameters
for both samples have been obtained from a very uniform spectroscopic analysis
\citep[][]{Santos-2004b}. Furthermore, the abundances of the three 
elements studied were derived from the analysis of high-resolution spectra
using the same line-lists for every star. The results show that planet
host stars have, on average, [X/H] values (with X = Na, Mg and Al) above 
those found in field dwarfs. However, no major
differences were found to exist between the [X/Fe] ratios 
in planet hosts and those found in our comparison sample stars for a given value
of [Fe/H].

\section{Data and spectral analysis}
\label{sec:data}

With the exception of \object{HD\,330075} \cite[][]{Pepe-2004}, the spectra used in this 
paper were those taken by \citet[][]{Santos-2004b} in their study of 98 planet-host stars 
and 41 comparison sample ``single'' dwarfs. For a more thorough presentation
of the samples see \citet[][and references therein]{Santos-2004b}. 
\object{HD\,219542\,B} was excluded
from the planet-host star list as the presence
of a planet around this star has been disproved \citep[][]{Desidera-2004}. 

The stellar parameters used are also the same has listed by \citet[][]{Santos-2004b} 
and \citet[][]{Pepe-2004}\footnote{Whose stellar
parameters have also been obtained by our team.}. These were
derived in a very uniform way, using high resolution spectra and the same line-lists and 
model atmospheres for all the stars.

\begin{table}[t!]
\caption{List of Na, Mg and Al lines and atomic parameters.}
\begin{tabular}{lccclcc}
\hline
$\lambda$ (\AA)& $\chi_l$ & $\log{gf}$ & & $\lambda$ (\AA)& $\chi_l$ & $\log{gf}$\\
\hline 
\multicolumn{3}{l}{Na ($\log{\epsilon}_\odot$=6.33)} & & 8736.02 & 5.946 & -0.224\\
5688.22 & 2.104 & -0.625                             & & 8923.57 & 5.394 & -1.652\\
6154.23 & 2.102 & -1.607                             & & \multicolumn{3}{l}{Al ($\log{\epsilon}_\odot$=6.47)} \\
6160.75 & 2.104 & -1.316                             & & 6696.03 & 3.143 & -1.570\\
\multicolumn{3}{l}{Mg ($\log{\epsilon}_\odot$=7.58)} & & 6698.67 & 3.143 & -1.879\\
5711.09 & 4.346 & -1.706                             & & 7835.31 & 4.022 & -0.728\\
6318.72 & 5.108 & -1.996                             & & 7836.13 & 4.022 & -0.559\\
6319.24 & 5.108 & -2.179                             & & 8772.87 & 4.022 & -0.425\\
8712.69 & 5.932 & -1.204                             & & 8773.91 & 4.022 & -0.212\\
\hline 
\end{tabular}
\label{tab:lines}
\end{table}

Abundances for the elements studied here were derived from the analysis of
a list of spectral lines initially taken from the literature \citep[][]{Edvardsson-1993a,Feltzing-1998,Gonzalez-2000,Gonzalez-2001,Smith-2001,
Sadakane-2002,Chen-2003}. A careful choice of these was done
based on the analysis of the Kurucz Solar Flux Atlas \citep[][]{Kurucz-1984} to include only those
lines that are not blended in the solar spectrum. All the lines used have wavelengths 
between 5000 and 9000\AA.

To derive semi-empirical atomic $\log{gf}$ values for the lines, we used their Equivalent Widths 
(EW) measured in the solar spectrum, and performed an 
inverted solar analysis. The solar abundances for each element were taken 
from \citet[][]{Anders-1989}, and the solar model was considered to
have T$_{\mathrm{eff}}$=5777\,K, $\log{g}$=4.44\,dex, $\xi_t$=1.00\,km\,s$^{-1}$ and 
$\log{\epsilon}$(Fe)=7.47\,dex. The final list of lines with their atomic parameters 
is presented in Table\,\ref{tab:lines}.

For each star, line EW were then measured in the stellar spectra 
using the IRAF ``splot'' task within the {\tt echelle} package. The 
analysis was done in Local Thermodynamic Equilibrium (LTE) using a revised 
version of the code MOOG \citep[][]{Sneden-1973} (with the {\tt abfind} 
driver) and a grid of \citep[][]{Kurucz-1993} ATLAS9 atmospheres. 
In a few particular cases, some lines were not used in the analysis due to the poor 
quality of the spectrum in the region of interest.

In Tables\,\ref{tab:planets1} through \ref{tab:comparison} we summarize the derived abundances for 
all stars with and without planetary-mass companions. 

\subsection{Errors}
\label{sec:errors}

\begin{figure}[t!]
\psfig{width=0.5\textwidth,file=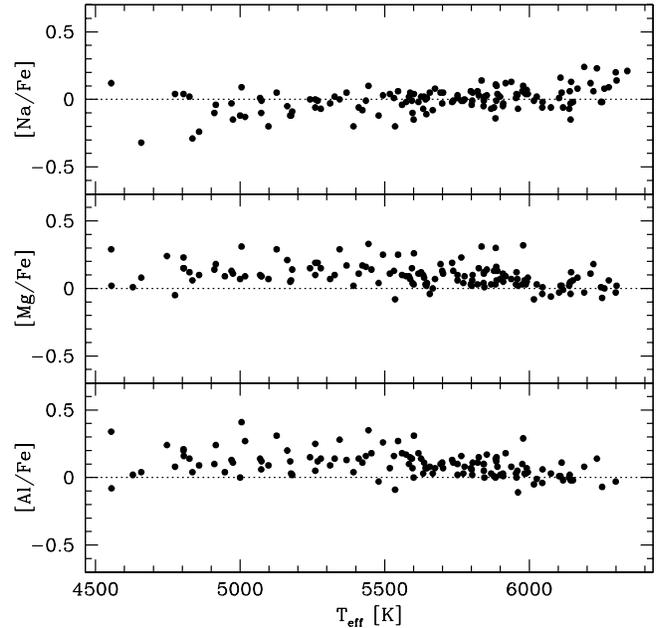}
\caption[]{[X/Fe] as a function of T$_{eff}$. The dotted line represents the solar [X/Fe] value.}
\label{fig:xteff}
\end{figure}

\begin{figure}[t!]
\psfig{width=0.5\textwidth,file=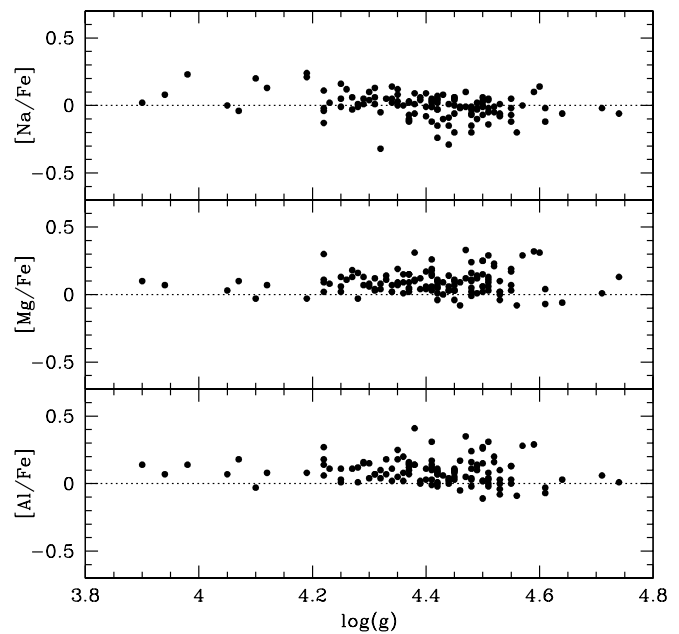}
\caption[]{Same as Fig.\,\ref{fig:xteff} but for $\log{g}$.}
\label{fig:xlogg}
\end{figure}

Errors and uncertainties can affect abundance measurements in various ways. Individual 
lines can be affected by errors in the EW, caused by unnoticed blends or a poor location 
of the continuum. Given that we have usually measured more than one line of each element
in a given star, the dispersion around the average abundance gives us an
idea about the errors due to these effects. Tables\,\ref{tab:planets1} 
through \ref{tab:comparison} show that the observed dispersions are usually below
0.05\,dex, attesting to the quality of the line-list used.

Table\,\ref{tab:sensitivity} shows the abundance sensitivity to changes in 
effective temperature, $\log{g}$, microturbulence and [Fe/H] in three of our targets.
A perturbation of 50 K in effective temperature, 0.15 dex in log $g$, 0.10 km\,s$^{-1}$ 
in microturbulence and 0.05 dex in [Fe/H] 
\citep[typical uncertainties on our parameters --][]{Santos-2004b} 
leads to a total typical uncertainty below 0.06 dex in the derived abundances.

Systematic errors in the stellar parameters and/or NLTE effects (the latter not taken 
into account in the current analysis) may also affect the derived 
abundances \citep[see e.g. ][]{Shi-2004}. 
In Figs.\,\ref{fig:xteff} and \ref{fig:xlogg} we present the abundance ratios
[Na/Fe], [Mg/Fe] and [Al/Fe] as functions of the effective temperature and surface 
gravity. The results show that a very small dependence seems to exist for the abundances 
as a function of T$_{\mathrm{eff}}$, in particular for [Na/Fe]. A fit to the 
points in Fig.\,\ref{fig:xteff}
shows, however, that the dependence is always below 0.10\,dex/1000\,K. \citet[][]{Shi-2004} 
have shown that the Na lines used in the current 
paper are particularly insensitive to NLTE effects.

\begin{table}[t!]
\caption{Abundance sensitivity to changes of 50 K in effective temperature, 
0.15 dex in log $g$, 0.10 km\,s$^{-1}$ in microturbulence and 0.10 dex in [Fe/H].}
\begin{tabular}{lcccc}
\hline
 & $\Delta$T$_{eff}$ & $\Delta$log $g$ & $\Delta\xi$ & $\Delta$[Fe/H] \\
 & +50 K & +0.15 dex & +0.10 km\,s$^{-1}$ & +0.10 dex \\
\hline 
\multicolumn{5}{c}{HD128311 (4835 K; 4.44 dex; 0.89 km\,s$^{-1}$; 0.03 dex)} \\
Na & 0.04 & -0.04 & -0.02 & 0.02 \\
Mg & 0.01 & -0.03 & -0.02 & 0.01 \\
Al & 0.03 & -0.01 & -0.02 & 0.00 \\
\multicolumn{5}{c}{HD 38529 (5674 K; 3.94 dex; 1.38 km\,s$^{-1}$; 0.40 dex)} \\
Na & 0.03 & -0.03 & -0.01 & -0.01 \\
Mg & 0.02 & -0.03 & -0.02 & 0.00 \\
Al & 0.02 & -0.01 & -0.01 & -0.01 \\
\multicolumn{5}{c}{HD 17051 (6252 K; 4.61 dex; 1.18 km\,s$^{-1}$; 0.26 dex)} \\
Na & 0.03 & -0.02 & 0.00 & 0.01 \\
Mg & 0.03 & -0.02 & 0.00 & 0.01 \\
Al & 0.03 & 0.00 & 0.00 & 0.00 \\
\hline
\end{tabular}    
\label{tab:sensitivity}
\end{table}

\section{Na, Mg and Al in planet-host stars}
\label{sec:abund}

\begin{figure*}[t]
\psfig{width=\textwidth,file=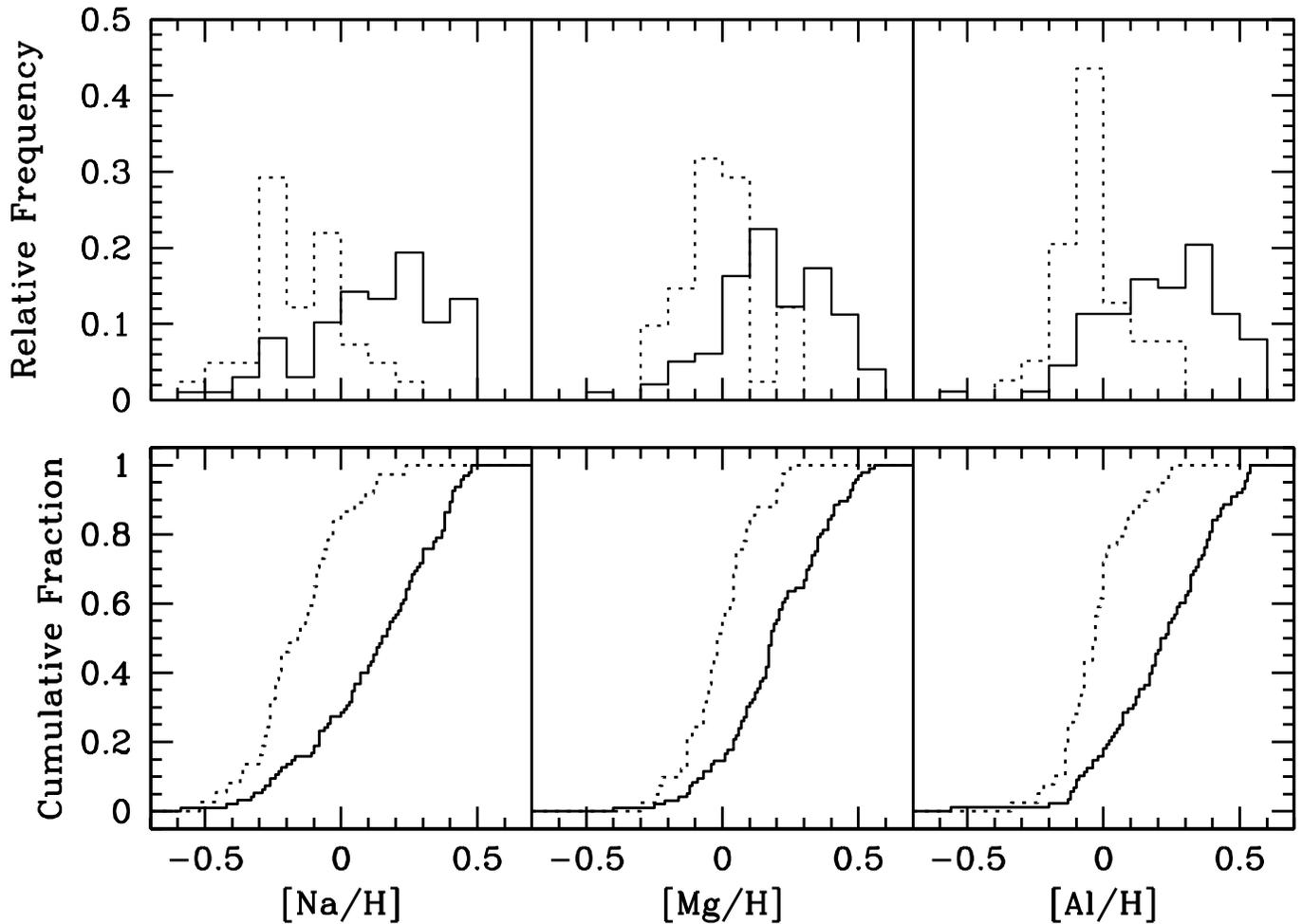}
\caption[]{{\it Upper panels}: distributions of [Na/H], [Mg/H] and [Al/H] for stars with planets 
(solid histogram) and for the comparison sample (dashed histogram). {\it Lower panels}: 
cumulative functions for both samples. In all cases, Kolmogorov-Smirnov probabilities 
are of the order of 10$^{-8}$ that the two samples belong to the same population.}
\label{fig:histo}
\end{figure*}

In Fig.\,\ref{fig:histo} we present the distributions of [X/H] (with X = Na, Mg and Al) 
for stars with planets 
and comparison stars, as well as their cumulative functions. These histograms are similar 
to those presented for [Fe/H] in \citet[][]{Santos-2004b}, and indicate that the 
excess metallicity for stars with planets is, as expected, not unique to iron. 
In Table\,\ref{tab:average}, we list the average abundances for each element 
in the two samples, as well as its mean standard deviation and the differences 
in abundances between stars with planets
and comparison stars. These differences vary between 0.20\,dex (Mg) and 0.28\,dex (Na),
and are comparable to those found for other metals \citep[][]{Bodaghee-2003,Ecuvillon-2004a,Ecuvillon-2004b,Santos-2004b}.

The [Na/H] and [Al/H] distributions for planet-host stars also reveal an interesting feature 
already discussed in the literature \citep[e.g.][]{Santos-2001,Bodaghee-2003} 
for other elements: the distributions are not symmetrical, increasing as a function
of increasing [X/H]. Possible interpretations 
for this are discussed in e.g. \citet[][]{Santos-2001,Santos-2004b}. The [Mg/H] distribution also
looks slightly bimodal. The reason for the lack of stars with [Mg/H]$\sim$0.3 is not 
clear, and may be simply due to small number statistics.

We have investigated possible dependences of the Na, Mg and Al abundances
with the parameters of the orbiting planets (period, mass, eccentricity). 
No statistically significant differences were found.

\begin{table}[t!]
\caption{Average abundances $<$[X/H]$>$ for stars with planets and for comparison 
sample stars. The rms around the mean and the abundance difference between the two 
samples for each element are also listed.}
\begin{tabular}{cccccc}
Element   & \multicolumn{2}{c}{Stars with planets}  & \multicolumn{2}{c}{Stars without planets} &  Average\\
$($X$)$          & $<$[X/H]$>$       & $\sigma$            & $<$[X/H]$>$    &    $\sigma$       & difference \\
\hline 
Na & 0.12 & 0.24 & -0.16 & 0.17 & 0.28 \\
Mg & 0.19 & 0.19 & -0.01 & 0.13 & 0.20 \\
Al & 0.21 & 0.21 & -0.03 & 0.13 & 0.24 \\
\hline
\end{tabular} 
\label{tab:average}
\end{table}

\subsection{[X/Fe] vs. [Fe/H]}
\label{subsec:xfe}

In \citet[][]{Gonzalez-2000} and \citet[][]{Gonzalez-2001} some possible anomalies concerning Na, Mg and Al 
were discussed. These authors have found that the [X/Fe] abundance ratios (with X=Na, Mg and Al) 
for stars with planets seemed to be slightly lower than those found for field 
dwarfs. Such a difference was, however, not found by other authors \citep[][]{Sadakane-2002}.

\begin{figure}[t!]
\psfig{width=0.5\textwidth,file=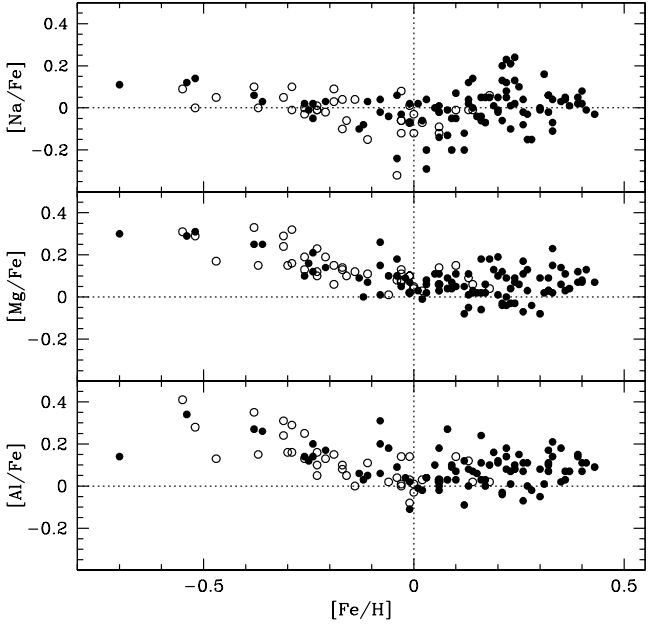}
\caption[]{Abundance ratios [X/Fe] for Na, Mg and Al as functions of metallicity. 
Filled circles are planet-host stars and the empty triangles are comparison stars. Dotted lines 
represent solar values.}
\label{fig:abund}
\end{figure}

\begin{figure}[t!]
\psfig{width=0.5\textwidth,file=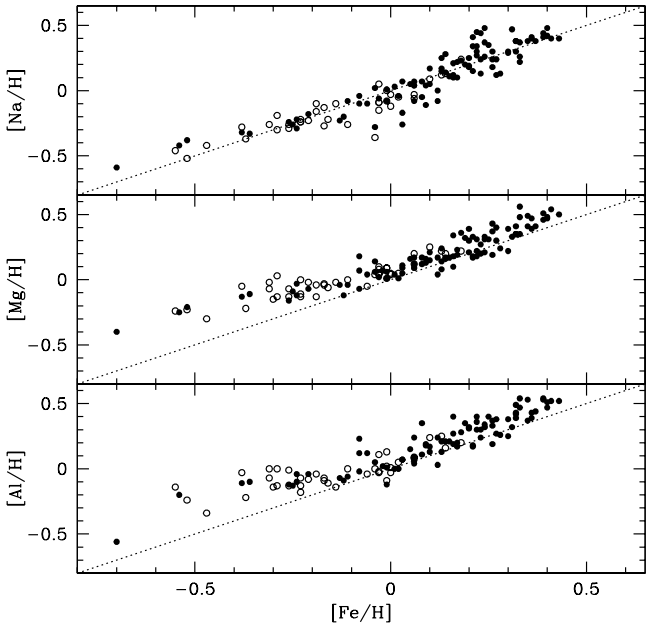}
\caption[]{Abundance ratios [X/H] for Na, Mg and Al as functions of [Fe/H]. 
Symbols as in Fig\,\ref{fig:abund}.}
\label{fig:xh}
\end{figure}

In Figs\,\ref{fig:abund} and \ref{fig:xh} we present plots of [X/Fe] and [X/H] vs. [Fe/H] for the
three elements. Filled circles represent planet-host stars, while open circles
denote ``single'' field dwarfs. As we can see from the plots, there is
no clear and distinctive difference between the two samples. 
We do not confirm that planet-host stars have lower [X/Fe] 
ratios for a given metallicity ([Fe/H]).
The abundance distributions
of stars with planets are high-[Fe/H] extensions to the curves traced by the field dwarfs,
and no discontinuity seems to exist.

We should caution, however, that the number of comparison stars plotted 
is quite small, in particular for the metal-rich domain. An extension of
the samples may thus be needed to support this discussion. Such a study is
currently in progress.

Aluminum and magnesium were shown to be good tracers 
of the thin- and thick-disk populations, the latter presenting higher 
[X/Fe] values for a given [Fe/H] \citep[e.g.][]{Bensby-2003,Fuhrmann-2004}. 
The fact that no clear difference exists between planet-hosts and single field dwarfs
suggests that both groups of objects belong (statistically) to the same 
population (the thin disk).

\section{Galactic chemical evolution trends}
\label{sec:galactic}

Although the main goal of this work is to compare the abundances of Na, Mg and Al 
for planet-hosts and stars without giant planets, the results presented also give
us a chance to study the chemical evolution trends of the Galaxy, and in 
particular in the metal-rich domain. 
Given the absence of significant differences between stars with 
planets and comparison stars, we will consider the samples as a whole
for the rest of the paper. 

\subsection{Sodium}
\label{subsec:sodium}

From Fig.\,\ref{fig:abund}, we can see that the distribution of [Na/Fe] presents a shallow
decrease as a function of [Fe/H] in the domain $-$0.7$<$[Fe/H]$<$0.0. 
For solar metallicity the average [Na/Fe] abundances are clearly below solar.
For [Fe/H]$>$0, we notice an uptrend. Interestingly, there 
is some hint that the [Na/Fe] values may become constant (and near the solar value) 
for [Fe/H] above $\sim$0.25\,dex. The average [Na/Fe] of the stars with [Fe/H]$>$0.0
is solar (0.00\,dex, with a dispersion of 0.10\,dex).
Unfortunately, the abundance distribution is affected 
by the dispersion of the points. It is not clear whether
this dispersion is real or an artifact of the analysis (see Sect.\,\ref{sec:errors}).

The global trend observed in this paper confirms the results presented 
by other authors \citep[e.g.][]{Edvardsson-1993a,Feltzing-1998,Reddy-2003,Bensby-2003,
Shi-2004}, with the exception of \citet[][]{Chen-2000}. 
The possible constance of the [Na/Fe] values 
for [Fe/H] above $\sim$0.25\,dex is also tentatively seen 
in Fig.\,13 of \citet[][]{Bensby-2003}, although it is not present
in the NLTE study of \citet[][]{Shi-2004}. 

Sodium, as well as aluminum, is though to be mostly a product of Ne and C burning 
in massive stars, the same objects that will later give rise to SN\,II, the major 
producers of Mg \citep[][]{Arnett-1996}. As we can see from Fig.\ref{fig:abund},
however, Na and Mg do not behave similarly as a function of stellar metallicity. This
difference, also discussed in \citet[][]{Bensby-2003}, may indicate the 
existence of different sources for one or both of these elements 
\citep[SN\,Ia and AGB stars -- ][]{Mowlavi-1999,Karakas-2003,Tsujimoto-1995}, 
a dependence of the SNe yields of both elements with stellar metallicity,
or a variation in the spectrum of the progenitor masses inside the metallicity
regime studied here. Interestingly, Na abundances increase (globaly) almost exactly as
iron (see Fig.\,\ref{fig:xh}), which may indicate a similar origin for both elements.

\subsection{Magnesium}
\label{subsec:magnesium}

As for Na, the [Mg/Fe] distribution decreases with increasing metallicity 
until [Fe/H]$\sim$0.0, although the decrease is stronger than for Na 
(the [Na/Fe] values always remain close to solar for [Fe/H]$<$0.0). 
For higher metallicities, however, the [Mg/Fe] values become constant at an average of 
0.06$\pm$0.06\,dex (where the error bar represents the rms), and no clear 
upturn in observed, although a slight increase of [Mg/Fe] as a function
of [Fe/H] cannot be excluded. The average [Mg/Fe] (for a given [Fe/H]) remains above
solar for the whole metallicity regime studied.

The Mg trend observed in Fig\,\ref{fig:abund} is very similar to the one
presented by other authors 
\citep[e.g.][]{Feltzing-1998,Chen-2000,Bensby-2003,Reddy-2003}.

Comparison of the Mg abundances thought to be produced by massive stars \citep[][]{Arnett-1996}
with the ones found for other alpha-elements like Si, Ca and Ti, also
produced by SN\,Ia events \citep[e.g.][]{Thielemann-2002}, should provide
evidence about the origin of these elements. If we compare the Mg abundances 
derived in the current paper with those for other alpha elements
presented in \citet[][]{Bodaghee-2003} (see their Fig.\,2), we see a global agreement 
with silicon, although for metallicities below solar, the [Si/Fe] abundances 
tend to be lower than those found for [Mg/Fe]. For 
calcium (presenting a constant downward trend as a function of [Fe/H]), and to 
a lesser extent for Ti, however, we do not find the same trends for [Fe/H]$>$0.0.
This may indicate that the yields for these two latter elements may be 
metallicity-dependent (e.g. due to a progenitor mass-metallicity dependence), 
or that different sources for these different elements give important contributions 
\citep[e.g. SN\,Ia and AGB stars -- ][]{Thielemann-2002,Karakas-2003}. 
Our result regarding this issue is similar to the one found by \citet[][]{Bensby-2003}. 

\subsection{Aluminum}
\label{subsec:aluminium}

The Al distribution is very similar to the one observed for Mg,
with the [Al/Fe] values decreasing with increasing [Fe/H] up to solar values,
then becoming relatively constant for metallicities above solar at an average
value of 0.07$\pm$0.07\,dex.
Contrary to Mg, however, for solar metallicities
the average [Al/Fe] of our stars is close to 0.0\,dex. 
Again, we cannot exclude a very shallow uptrend for above-solar metallicities.

Globally, the trend observed for Al is similar to the one already presented by
other authors \citep[e.g.][]{Edvardsson-1993a,Feltzing-1998,Reddy-2003}, although the slight
upturn seen in Fig.\,13 of \citet[][]{Bensby-2003} for [Fe/H] above solar is not
clear in our data. Also, as for Na, our results are considerably different from those
obtained by \citet[][]{Chen-2000}.

The very similar trend observed for Al and Mg suggests a similar origin for both
elements.

\subsection{Anomalies and planets}

In Sect.\,\ref{subsec:xfe} we have seen that no large differences in the [X/Fe] ratios 
were found between planet-hosts and single stars. However, it is still
possible that some specific trends found for the metal-rich tail
of the Galactic chemical evolution are due to the presence of planets.
Indeed, planets are more frequent around metal-rich stars \citep[e.g.][]{Santos-2004b}.
It should thus be no surprise if their presence influences the
yields of certain elements for metal-rich stars in the AGB phase
responsible for the chemical enrichment of the Galaxy.

It is known that in addition to the CNO elements, Na, Mg and Al abundances are
correlated in a large number of globular cluster stars \citep[][]{Kraft-1994} showing
the general effects of mixing. It also became clear that Na and Al 
enhancements observed in many globular-cluster giants \citep[][]{Kraft-1994} can be 
produced by the $^{22}$Ne(p,$\gamma$)$^{23}$Na reaction of the NeNa-cycle 
\citep[][and references therein]{Langer-1993}. For this to occur, one would need 
material that had undergone significant O$\rightarrow$N or C$\rightarrow$N processing, 
while in the O-depleted layers we could expect a transformation of the $^{26}$Mg and
$^{25}$Mg isotopes into Al in the MgAl-cycle \citep[][]{Langer-1993}. 
\citet[][]{Langer-1997} have also shown that the use of
updated rates of Ne-Na cycle reactions can explain the Na-O abundance anticorrelation observed in 
globular-cluster giants, and the Na-N correlation
observed in field halo giants. 
It is thus apparent that giants are not reliable
for studying the primordial abundances of CNO-cycle elements. Clearly, it is
not safe to use CNO, Na, Al and Mg elements in giants as probes of Galactic
chemical evolution since rotation and mixing can alter surface abundances of these
elements. 

Furthermore, it is not clear how the yields of these elements will be altered in 
metal-rich giants due to the presence of planets. 
As a tentative example, if a star during its AGB evolution engulfs 
(short period) planets (many such planets have been found by current radial-velocity 
surveys\footnote{See e.g. table at http://obswww.unige.ch/exoplanets}), 
it will change its angular momentum and rotation. As a consequence, its internal mixing
history may be changed, possibly bringing different quantities of produced elements
to the stellar outer layers, thus changing the chemical yields. Planet 
engulfment, eventually followed by extra-mixing, has already been suggested 
to explain the existence of Li-rich giant stars \citep[e.g.][]{Alexander-1967,Siess-1999}.
Overall, these processes could produce important changes in the chemical evolution trends
of the Galaxy, in particular for the metal-rich domain.

\section{Concluding remarks}
\label{sec:remarks}

In this paper we have derived abundances of Na, Mg and Al in a set of 98 planet-host stars 
and 41 comparison stars without known planets. The results were derived through a detailed 
and uniform spectroscopic analysis and gave us the possibility to compare the two samples. 
The main goal of this comparison is to look for differences connected to the presence 
of giant planets, but it was also used to explore chemical evolution trends in the Galaxy.

Through a comparison of Na, Mg and Al abundances of planet-hosts and stars without planets, we 
found an overabundance similar to that already found in iron. However, the abundance ratios [X/Fe] 
(with X=Na, Mg and Al) found for planet-hosts were not found to be significantly different 
from those obtained for our comparison stars for a given [Fe/H]. 
Finally, the trends in the [X/Fe] vs. [Fe/H] plots derived in this paper are
similar to the ones presented in the literature, although some new interesting
details seem to appear in the metal-rich domain. A discussion of possible implications
for the chemical evolution of the Galaxy is presented.

Given the lack of ``single'' comparison stars with [Fe/H]$>$0.2\,dex, the
conclusions presented above must be seen as preliminary. 
Indeed, future studies of elemental abundances in planet-host stars
should include a comparison with a more significant number of ``single'' 
dwarfs having metallicities above solar. Such work is currently in progress.

\begin{acknowledgements}
  Support from Funda\c{c}\~ao para a Ci\^encia e Tecnologia (Portugal) 
  to N.C.S. in the form of a scholarship is gratefully acknowledged.
  We wish to thank the Swiss National Science Foundation (Swiss NSF) 
  for the continuous support for this project.
\end{acknowledgements}

\bibliographystyle{aa}
\bibliography{santos_bibliography}

\begin{thebibliography}{46}
\expandafter\ifx\csname natexlab\endcsname\relax\def\natexlab#1{#1}\fi

\bibitem[{{Alexander}(1967)}]{Alexander-1967}
{Alexander}, J.~B. 1967, The Observatory, 87, 238

\bibitem[{{Anders} \& {Grevesse}(1989)}]{Anders-1989}
{Anders}, E. \& {Grevesse}, N. 1989, Geochimica et Cosmochimica Acta, 53, 197

\bibitem[{{Arnett}(1996)}]{Arnett-1996}
{Arnett}, D. 1996, {Supernovae and nucleosynthesis. an investigation of the
  history of matter, from the Big Bang to the present} (Princeton series in
  astrophysics, Princeton, NJ: Princeton University Press, |c1996)

\bibitem[{{Bensby} {et~al.}(2003){Bensby}, {Feltzing}, \& {Lundstr{\"
  o}m}}]{Bensby-2003}
{Bensby}, T., {Feltzing}, S., \& {Lundstr{\" o}m}, I. 2003, A\&A, 410, 527

\bibitem[{{Bodaghee} {et~al.}(2003){Bodaghee}, {Santos}, {Israelian}, \&
  {Mayor}}]{Bodaghee-2003}
{Bodaghee}, A., {Santos}, N.~C., {Israelian}, G., \& {Mayor}, M. 2003, A\&A,
  404, 715

\bibitem[{{Chen} {et~al.}(2000){Chen}, {Nissen}, {Zhao}, {Zhang}, \&
  {Benoni}}]{Chen-2000}
{Chen}, Y.~Q., {Nissen}, P.~E., {Zhao}, G., {Zhang}, H.~W., \& {Benoni}, T.
  2000, A\&AS, 141, 491

\bibitem[{{Chen} {et~al.}(2003){Chen}, {Zhao}, {Nissen}, {Bai}, \&
  {Qiu}}]{Chen-2003}
{Chen}, Y.~Q., {Zhao}, G., {Nissen}, P.~E., {Bai}, G.~S., \& {Qiu}, H.~M. 2003,
  ApJ, 591, 925

\bibitem[{{Deliyannis} {et~al.}(2000){Deliyannis}, {Cunha}, {King}, \&
  {Boesgaard}}]{Deliyannis-2000}
{Deliyannis}, C.~P., {Cunha}, K., {King}, J.~R., \& {Boesgaard}, A.~M. 2000,
  AJ, 119, 2437

\bibitem[{{Desidera} {et~al.}(2004){Desidera}, {Gratton}, {Endl}, {Claudi}, \&
  {Cosentino}}]{Desidera-2004}
{Desidera}, S., {Gratton}, R.~G., {Endl}, M., {Claudi}, R.~U., \& {Cosentino},
  R. 2004, A\&A, 420, L27

\bibitem[{{Ecuvillon} {et~al.}(2004{\natexlab{a}}){Ecuvillon}, {Israelian},
  {Santos}, {Mayor}, {Garc{\'{\i}}a L{\' o}pez}, \&
  {Randich}}]{Ecuvillon-2004a}
{Ecuvillon}, A., {Israelian}, G., {Santos}, N.~C., {et~al.} 2004{\natexlab{a}},
  A\&A, 418, 703

\bibitem[{{Ecuvillon} {et~al.}(2004{\natexlab{b}}){Ecuvillon}, {Israelian},
  {Santos}, {Mayor}, {Villar}, \& {Bihain}}]{Ecuvillon-2004b}
{Ecuvillon}, A., {Israelian}, G., {Santos}, N.~C., {et~al.} 2004{\natexlab{b}},
  A\&A, 426, 619

\bibitem[{{Edvardsson} {et~al.}(1993){Edvardsson}, {Andersen}, {Gustafsson},
  {Lambert}, {Nissen}, \& {Tomkin}}]{Edvardsson-1993a}
{Edvardsson}, B., {Andersen}, J., {Gustafsson}, B., {et~al.} 1993, A\&A, 275,
  101

\bibitem[{{Feltzing} \& {Gustafsson}(1998)}]{Feltzing-1998}
{Feltzing}, S. \& {Gustafsson}, B. 1998, A\&AS, 129, 237

\bibitem[{{Fuhrmann}(2004)}]{Fuhrmann-2004}
{Fuhrmann}, K. 2004, Astronomische Nachrichten, 325, 3

\bibitem[{{Garcia Lopez} \& {Perez de Taoro}(1998)}]{GarciaLopez-1998}
{Garcia Lopez}, R.~J. \& {Perez de Taoro}, M.~R. 1998, A\&A, 334, 599

\bibitem[{{Gonzalez}(1998)}]{Gonzalez-1998}
{Gonzalez}, G. 1998, A\&A, 334, 221

\bibitem[{{Gonzalez} \& {Laws}(2000)}]{Gonzalez-2000}
{Gonzalez}, G. \& {Laws}, C. 2000, AJ, 119, 390

\bibitem[{{Gonzalez} {et~al.}(2001){Gonzalez}, {Laws}, {Tyagi}, \&
  {Reddy}}]{Gonzalez-2001}
{Gonzalez}, G., {Laws}, C., {Tyagi}, S., \& {Reddy}, B.~E. 2001, AJ, 121, 432

\bibitem[{{Israelian} {et~al.}(2003){Israelian}, {Santos}, {Mayor}, \&
  {Rebolo}}]{Israelian-2003}
{Israelian}, G., {Santos}, N.~C., {Mayor}, M., \& {Rebolo}, R. 2003, A\&A, 405,
  753

\bibitem[{{Israelian} {et~al.}(2004){Israelian}, {Santos}, {Mayor}, \&
  {Rebolo}}]{Israelian-2004}
{Israelian}, G., {Santos}, N.~C., {Mayor}, M., \& {Rebolo}, R. 2004, A\&A, 414,
  601

\bibitem[{{Karakas} \& {Lattanzio}(2003)}]{Karakas-2003}
{Karakas}, A.~I. \& {Lattanzio}, J.~C. 2003, PASA, 20, 279

\bibitem[{{Kraft}(1994)}]{Kraft-1994}
{Kraft}, R.~P. 1994, PASP, 106, 553

\bibitem[{{Kurucz}(1993)}]{Kurucz-1993}
{Kurucz}, R. 1993, ATLAS9 Stellar Atmosphere Programs and 2 km/s grid.~Kurucz
  CD-ROM No.~13.~ Cambridge, Mass.: Smithsonian Astrophysical Observatory,
  1993., 13

\bibitem[{{Kurucz} {et~al.}(1984){Kurucz}, {Furenlid}, \&
  {Brault}}]{Kurucz-1984}
{Kurucz}, R.~L., {Furenlid}, I., \& {Brault}, J.~T.~L. 1984, {Solar flux atlas
  from 296 to 1300 nm} (National Solar Observatory Atlas, Sunspot, New Mexico:
  National Solar Observatory, 1984)

\bibitem[{{Langer} {et~al.}(1993){Langer}, {Hoffman}, \&
  {Sneden}}]{Langer-1993}
{Langer}, G.~E., {Hoffman}, R., \& {Sneden}, C. 1993, PASP, 105, 301

\bibitem[{{Langer} {et~al.}(1997){Langer}, {Hoffman}, \&
  {Zaidins}}]{Langer-1997}
{Langer}, G.~E., {Hoffman}, R.~E., \& {Zaidins}, C.~S. 1997, PASP, 109, 244

\bibitem[{{Mowlavi}(1999)}]{Mowlavi-1999}
{Mowlavi}, N. 1999, A\&A, 350, 73

\bibitem[{{Pepe} {et~al.}(2004){Pepe}, {Mayor}, {Queloz}, {Benz}, {Bonfils},
  {Bouchy}, {Curto}, {Lovis}, {M{\' e}gevand}, {Moutou}, {Naef}, {Rupprecht},
  {Santos}, {Sivan}, {Sosnowska}, \& {Udry}}]{Pepe-2004}
{Pepe}, F., {Mayor}, M., {Queloz}, D., {et~al.} 2004, A\&A, 423, 385

\bibitem[{{Reddy} {et~al.}(2003){Reddy}, {Tomkin}, {Lambert}, \& {Allende
  Prieto}}]{Reddy-2003}
{Reddy}, B.~E., {Tomkin}, J., {Lambert}, D.~L., \& {Allende Prieto}, C. 2003,
  MNRAS, 340, 304

\bibitem[{{Reid}(2002)}]{Reid-2002}
{Reid}, I.~N. 2002, PASP, 114, 306

\bibitem[{{Ryan}(2000)}]{Ryan-2000}
{Ryan}, S.~G. 2000, MNRAS, 316, L35

\bibitem[{{Sadakane} {et~al.}(2002){Sadakane}, {Ohkubo}, {Takeda}, {Sato},
  {Kambe}, \& {Aoki}}]{Sadakane-2002}
{Sadakane}, K., {Ohkubo}, M., {Takeda}, Y., {et~al.} 2002, PASJ, 54, 911

\bibitem[{{Santos} {et~al.}(2004{\natexlab{a}}){Santos}, {Bouchy}, {Mayor},
  {Pepe}, {Queloz}, {Udry}, {Lovis}, {Bazot}, {Benz}, {Bertaux}, {Lo Curto},
  {Delfosse}, {Mordasini}, {Naef}, {Sivan}, \& {Vauclair}}]{Santos-2004a}
{Santos}, N.~C., {Bouchy}, F., {Mayor}, M., {et~al.} 2004{\natexlab{a}}, A\&A,
  426, L19

\bibitem[{{Santos} {et~al.}(2002){Santos}, {Garc{\'{\i}}a L{\' o}pez},
  {Israelian}, {Mayor}, {Rebolo}, {Garc{\'{\i}}a-Gil}, {P{\' e}rez de Taoro},
  \& {Randich}}]{Santos-2002c}
{Santos}, N.~C., {Garc{\'{\i}}a L{\' o}pez}, R.~J., {Israelian}, G., {et~al.}
  2002, A\&A, 386, 1028

\bibitem[{{Santos} {et~al.}(2004{\natexlab{b}}){Santos}, {Israelian},
  {Garc{\'{\i}}a L{\' o}pez}, {Mayor}, {Rebolo}, {Randich}, {Ecuvillon}, \&
  {Dom{\'{\i}}nguez Cerde{\~ n}a}}]{Santos-2004c}
{Santos}, N.~C., {Israelian}, G., {Garc{\'{\i}}a L{\' o}pez}, R.~J., {et~al.}
  2004{\natexlab{b}}, A\&A, 427, 1085

\bibitem[{{Santos} {et~al.}(2000){Santos}, {Israelian}, \&
  {Mayor}}]{Santos-2000b}
{Santos}, N.~C., {Israelian}, G., \& {Mayor}, M. 2000, A\&A, 363, 228

\bibitem[{{Santos} {et~al.}(2001){Santos}, {Israelian}, \&
  {Mayor}}]{Santos-2001}
{Santos}, N.~C., {Israelian}, G., \& {Mayor}, M. 2001, A\&A, 373, 1019

\bibitem[{{Santos} {et~al.}(2004{\natexlab{c}}){Santos}, {Israelian}, \&
  {Mayor}}]{Santos-2004b}
{Santos}, N.~C., {Israelian}, G., \& {Mayor}, M. 2004{\natexlab{c}}, A\&A, 415,
  1153

\bibitem[{{Santos} {et~al.}(2003){Santos}, {Israelian}, {Mayor}, {Rebolo}, \&
  {Udry}}]{Santos-2003}
{Santos}, N.~C., {Israelian}, G., {Mayor}, M., {Rebolo}, R., \& {Udry}, S.
  2003, A\&A, 398, 363

\bibitem[{{Shi} {et~al.}(2004){Shi}, {Gehren}, \& {Zhao}}]{Shi-2004}
{Shi}, J.~R., {Gehren}, T., \& {Zhao}, G. 2004, A\&A, 423, 683

\bibitem[{{Siess} \& {Livio}(1999)}]{Siess-1999}
{Siess}, L. \& {Livio}, M. 1999, MNRAS, 308, 1133

\bibitem[{{Smith} {et~al.}(2001){Smith}, {Cunha}, \& {Lazzaro}}]{Smith-2001}
{Smith}, V.~V., {Cunha}, K., \& {Lazzaro}, D. 2001, AJ, 121, 3207

\bibitem[{{Sneden}(1973)}]{Sneden-1973}
{Sneden}, C. 1973, Ph.D. Thesis, Univ. of Texas

\bibitem[{{Takeda} {et~al.}(2001){Takeda}, {Sato}, {Kambe}, {Aoki}, {Honda},
  {Kawanomoto}, {Masuda}, {Izumiura}, {Watanabe}, {Koyano}, {Maehara},
  {Norimoto}, {Okuda}, {Shimizu}, {Uraguchi}, {Yanagisawa}, {Yoshida},
  {Miyama}, \& {Ando}}]{Takeda-2001}
{Takeda}, Y., {Sato}, B., {Kambe}, E., {et~al.} 2001, PASJ, 53, 1211

\bibitem[{{Thielemann} {et~al.}(2002){Thielemann}, {Argast}, {Brachwitz},
  {Martinez-Pinedo}, {Rauscher}, {Liebend{\" o}rfer}, {Mezzacappa}, {H{\"
  o}flich}, \& {Nomoto}}]{Thielemann-2002}
{Thielemann}, F.-K., {Argast}, D., {Brachwitz}, F., {et~al.} 2002, A\&SS, 281,
  25

\bibitem[{{Tsujimoto} {et~al.}(1995){Tsujimoto}, {Nomoto}, {Yoshii},
  {Hashimoto}, {Yanagida}, \& {Thielemann}}]{Tsujimoto-1995}
{Tsujimoto}, T., {Nomoto}, K., {Yoshii}, Y., {et~al.} 1995, MNRAS, 277, 945

\end{thebibliography}

\def\baselinestretch{1}

\begin{table*}[t!]
\center
\caption[]{Atmospheric parameters \citep[from ][]{Santos-2004b} and abundances of 
Na, Mg \& Al for stars with planets from HD142 to HD95128. The number of spectral lines used is 
given by $n$ while $\sigma$ denotes the rms around the average.}
\begin{tabular}{lccccccccccccc}
\hline
Star & [Na/H] & $\sigma$ & $n$ & [Mg/H] & $\sigma$ & $n$ & [Al/H] & $\sigma$ & $n$ & [Fe/H] & T$_{eff}$ & log$g$ & $\xi_t$\\
\hline
\object{HD\,142}  &	0.28&	0.06&	2&	0.16&	0.03&	2&	n.d.&	n.d.&	0&	0.14&	6302&	4.34&	1.86\\
\object{HD\,1237} &	-0.08&	0.06&	3&	0.04&	0.03&	3&	0.03&	0.06&	2&	0.12&	5536&	4.56&	1.33\\
\object{HD\,2039} &	0.38&	0.05&	2&	0.35&	0.03&	3&	0.42&	0.02&	2&	0.32&	5976&	4.45&	1.26\\
\object{HD\,3651} &	0.00&	0.00&	1&	0.17&	0.05&	3&	0.24&	0.03&	4&	0.12&	5173&	4.37&	0.74\\
\object{HD\,4203} &	  0.42&   0.02&   3&	  0.48&   0.05&   3&	  0.51&   0.02&   2&	  0.40&   5636&   4.23&   1.12\\
\object{HD\,4208} &	  -0.22&  0.03&   3&	  -0.12&  0.08&   3&	  -0.10&  0.01&   2&	  -0.24&  5626&   4.49&   0.95\\
\object{HD\,6434} &	  -0.38&  0.06&   2&	  -0.21&  0.06&   3&	  n.d.&   n.d.&   0&	  -0.52&  5835&   4.60&   1.53\\
\object{HD\,8574} &	0.04&	0.03&	3&	0.12&	0.06&	3&	0.04&	0.06&	2&	0.06&	6151&	4.51&	1.45\\
\object{HD\,9826} &	0.25&	0.02&	2&	0.24&	0.05&	3&	n.d.&	n.d.&	0&	0.13&	6212&	4.26&	1.69\\
\object{HD\,10647} &	  -0.06&  0.06&   3&	  0.02&   0.06&   2&	  n.d.&   n.d.&   0&	  -0.03&  6143&   4.48&   1.40\\
\object{HD\,10697} &	  0.14&   0.02&   3&	  0.17&   0.06&   3&	  0.21&   0.03&   2&	  0.14&   5641&   4.05&   1.13\\
\object{HD\,12661} &	  0.41&   0.06&   3&	  0.47&   0.03&   3&	  0.43&   0.02&   4&	  0.36&   5702&   4.33&   1.05\\
\object{HD\,13445} &	  -0.29&  0.06&   3&	  -0.03&  0.03&   3&	  -0.04&  0.01&   2&	  -0.24&  5163&   4.52&   0.72\\
\object{HD\,16141} &	  0.11&   0.03&   3&	  0.17&   0.02&   3&	  0.21&   0.03&   2&	  0.15&   5801&   4.22&   1.34\\
\object{HD\,17051} &	  0.24&   0.01&   3&	  0.19&   0.04&   3&	  0.19&   0.01&   2&	  0.26&   6252&   4.61&   1.18\\
\object{HD\,19994} &	  0.48&   0.02&   3&	  0.21&   0.00&   1&	  0.32&   0.00&   1&	  0.24&   6190&   4.19&   1.54\\
\object{HD\,20367} &	  0.10&   0.01&   3&	  0.19&   0.06&   3&	  0.17&   0.01&   2&	  0.17&   6138&   4.53&   1.22\\
\object{HD\,22049} &	  -0.23&  0.04&   3&	  -0.04&  0.04&   3&	  -0.07&  0.01&   2&	  -0.13&  5073&   4.43&   1.05\\
\object{HD\,23079} &	  -0.08&  0.03&   3&	  -0.04&  0.05&   3&	  -0.06&  0.02&   2&	  -0.11&  5959&   4.35&   1.20\\
\object{HD\,23596} &	  0.47&   0.05&   3&	  0.33&   0.05&   3&	  0.32&   0.04&   4&	  0.31&   6108&   4.25&   1.30\\
\object{HD\,27442} &	  0.41&   0.06&   2&	  0.51&   0.04&   2&	  0.53&   0.02&   2&	  0.39&   4825&   3.55&   1.18\\
\object{HD\,28185} &	  0.27&   0.04&   3&	  0.18&   0.04&   3&	  0.30&   0.02&   2&	  0.22&   5656&   4.45&   1.01\\
\object{HD\,30177} &	  0.44&   0.03&   3&	  0.46&   0.05&   4&	  0.54&   0.03&   5&	  0.39&   5588&   4.29&   1.08\\
\object{HD\,33636} &	  -0.10&  0.04&   3&	  -0.07&  0.01&   3&	  -0.02&  0.01&   2&	  -0.08&  6046&   4.71&   1.79\\
\object{HD\,37124} &	  -0.32&  0.05&   3&	  -0.13&  0.02&   3&	  -0.11&  0.02&   4&	  -0.38&  5546&   4.50&   0.80\\
\object{HD\,38529} &	  0.48&   0.01&   3&	  0.47&   0.02&   3&	  0.47&   0.03&   2&	  0.40&   5674&   3.94&   1.38\\
\object{HD\,39091} &	  0.17&   0.03&   3&	  0.15&   0.03&   3&	  0.17&   0.03&   2&	  0.10&   5991&   4.42&   1.24\\
\object{HD\,40979} &	  0.34&   0.00&   1&	  0.33&   0.04&   2&	  n.d.&   n.d.&   0&	  0.21&   6145&   4.31&   1.29\\
\object{HD\,46375} &	  0.19&   0.03&   2&	  0.39&   0.02&   3&	  0.32&   0.05&   4&	  0.20&   5268&   4.41&   0.97\\
\object{HD\,47536} &	  -0.42&  0.03&   3&	  -0.25&  0.04&   6&	  -0.20&  0.07&   6&	  -0.54&  4554&   2.48&   1.82\\
\object{HD\,49674} &	  0.22&   0.04&   2&	  0.35&   0.06&   3&	  n.d.&   n.d.&   0&	  0.33&   5644&   4.37&   0.89\\
\object{HD\,50554} &	  0.03&   0.05&   3&	  0.04&   0.02&   3&	  0.00&   0.01&   2&	  0.01&   6026&   4.41&   1.11\\
\object{HD\,52265} &	  0.24&   0.04&   3&	  0.20&   0.04&   3&	  0.24&   0.00&   1&	  0.23&   6103&   4.28&   1.36\\
\object{HD\,65216} &	  -0.20&  0.00&   3&	  -0.12&  0.03&   4&	  -0.09&  0.05&   6&	  -0.12&  5666&   4.53&   1.06\\
\object{HD\,68988} &	  0.40&   0.00&   1&	  0.39&   0.02&   3&	  0.39&   0.04&   4&	  0.36&   5988&   4.45&   1.25\\
\object{HD\,70642} &	  0.23&   0.05&   2&	  0.36&   0.04&   3&	  0.28&   0.02&   2&	  0.18&   5693&   4.41&   1.01\\
\object{HD\,72659} &	  0.07&   0.02&   3&	  0.11&   0.05&   4&	  0.07&   0.05&   6&	  0.03&   5995&   4.30&   1.42\\
\object{HD\,73256} &	  0.30&   0.05&   3&	  0.37&   0.03&   4&	  0.33&   0.05&   6&	  0.26&   5518&   4.42&   1.22\\
\object{HD\,73526} &	  0.24&   0.00&   2&	  0.40&   0.04&   4&	  0.38&   0.05&   6&	  0.27&   5699&   4.27&   1.26\\
\object{HD\,74156} &	  0.21&   0.01&   3&	  0.18&   0.01&   3&	  0.27&   0.00&   1&	  0.16&   6112&   4.34&   1.38\\
\object{HD\,75289} &	  0.13&   0.00&   2&	  0.24&   0.01&   2&	  0.26&   0.01&   2&	  0.28&   6143&   4.42&   1.53\\
\object{HD\,75732} &	  0.26&   0.03&   2&	  0.48&   0.05&   3&	  0.47&   0.01&   2&	  0.33&   5279&   4.37&   0.98\\
\object{HD\,76700} &	  0.40&   0.04&   3&	  0.54&   0.04&   6&	  0.52&   0.03&   6&	  0.41&   5737&   4.25&   1.18\\
\object{HD\,80606} &	  0.30&   0.05&   2&	  0.41&   0.04&   3&	  0.49&   0.03&   3&	  0.32&   5574&   4.46&   1.14\\
\object{HD\,82943} &	  0.29&   0.01&   2&	  0.22&   0.03&   2&	  0.25&   0.04&   2&	  0.30&   6016&   4.46&   1.13\\
\object{HD\,83443} &	  n.d.&   0.00&   0&	  0.49&   0.03&   3&	  0.53&   0.04&   2&	  0.35&   5454&   4.33&   1.08\\
\object{HD\,89744} &	  0.45&   0.02&   2&	  n.d.&   n.d.&   0&	  0.36&   0.05&   5&	  0.22&   6234&   3.98&   1.62\\
\object{HD\,92788} &	  0.38&   0.06&   3&	  0.35&   0.01&   3&	  0.43&   0.02&   2&	  0.32&   5821&   4.45&   1.16\\
\object{HD\,95128} &	  0.07&   0.02&   2&	  0.09&   0.04&   3&	  0.09&   0.03&   4&	  0.06&   5954&   4.44&   1.30\\
\hline
\end{tabular}
\label{tab:planets1}										     
\end{table*}

\newpage

\begin{table*}[t!]
\center
\caption{Same as Table\,\ref{tab:planets1} for stars with planets from HD\,106252 to HD\,330075.}
\begin{tabular}{lccccccccccccc}
\hline
Star & [Na/H] & $\sigma$ & $n$ & [Mg/H] & $\sigma$ & $n$ & [Al/H] & $\sigma$ & $n$ & [Fe/H] & T$_{eff}$ & log$g$ & $\xi_t$\\
\hline
\object{HD\,106252} &0.01    &0.01    &3     &  0.06   & 0.05  &  2   &    0.01   & 0.00  &  1&       -0.01  & 5899  &  4.34  &  1.08\\
\object{HD\,108147} &0.18    &0.05    &3     &  0.21   & 0.01  &  2   &    n.d.   & n.d.  &  0&       0.20   & 6248  &  4.49  &  1.35\\
\object{HD\,108874} &0.13    &0.06    &3     &  0.27   & 0.02  &  5   &    0.30   & 0.03  &  5&       0.23   & 5596  &  4.37  &  0.89\\
\object{HD\,111232} &-0.33   &0.05    &3     &  -0.11  & 0.02  &  3   &    -0.1   & 0.04  &  6&       -0.36  & 5494  &  4.50  &  0.84\\
\object{HD\,114386} &n.d.    &n.d.    &0     &  0.07   & 0.05  &  3   &    0.12   & 0.01  &  2&       -0.08  & 4804  &  4.36  &  0.57\\
\object{HD\,114729} &-0.26   &0.04    &3     &  -0.09  & 0.04  &  4   &    -0.13  & 0.05  &  4&       -0.25  & 5886  &  4.28  &  1.25\\
\object{HD\,114762} &-0.59   &0.02    &2     &  -0.40  & 0.00  &  2   &    -0.56  & 0.02  &  2&       -0.70  & 5884  &  4.22  &  1.31\\
\object{HD\,114783} &-0.11   &0.01    &2     &  0.16   & 0.03  &  3   &    0.18   & 0.05  &  4&       0.09   & 5098  &  4.45  &  0.74\\
\object{HD\,117176} &-0.10   &0.02    &3     &  0.04   & 0.03  &  3   &    0.12   & 0.06  &  2&       -0.06  & 5560  &  4.07  &  1.18\\
\object{HD\,120136} &0.44    &0.00    &1     &  n.d.   & n.d.  &  0   &    n.d.   & n.d.  &  0&       0.23   & 6339  &  4.19  &  1.70 \\
\object{HD\,121504} &0.10    &0.02    &3     &  0.10   & 0.05  &  3   &    0.19   & 0.06  &  2&       0.16   & 6075  &  4.64  &  1.31\\
\object{HD\,128311} &-0.26   &0.04    &2     &  0.09   & 0.05  &  5   &    0.07   & 0.03  &  4&       0.03   & 4835  &  4.44  &  0.89\\
\object{HD\,130322} &-0.17   &0.06    &2     &  0.05   & 0.06  &  3   &    0.07   & 0.00  &  1&       0.03   & 5392  &  4.48  &  0.85\\
\object{HD\,134987} &0.30    &0.04    &2     &  0.39   & 0.04  &  3   &    0.38   & 0.03  &  3&       0.30   & 5776  &  4.36  &  1.09\\
\object{HD\,136118} &0.02    &0.00    &1     &  0.14   & 0.01  &  2   &    n.d.   & n.d.  &  0&       -0.04  & 6222  &  4.27  &  1.79\\
\object{HD\,137759} &0.17    &0.07    &2     &  0.08   & 0.04  &  3   &    0.21   & 0.02  &  2&       0.13   & 4775  &  3.09  &  1.78\\
\object{HD\,141937} &0.05    &0.02    &3     &  0.21   & 0.04  &  3   &    0.13   & 0.03  &  4&       0.10   & 5909  &  4.51  &  1.13\\
\object{HD\,142415} &0.15    &0.03    &3     &  0.17   & 0.03  &  5   &    0.17   & 0.03  &  6&       0.21   & 6045  &  4.53  &  1.12\\
\object{HD\,143761} &-0.18   &0.00    &1     &  -0.07  & 0.01  &  3   &    -0.04  & 0.05  &  4&       -0.21  & 5853  &  4.41  &  1.35\\
\object{HD\,145675} &0.40    &0.03    &2     &  0.50   & 0.04  &  3   &    0.52   & 0.03  &  2&       0.43   & 5311  &  4.42  &  0.92\\
\object{HD\,147513} &-0.08   &0.02    &3     &  0.09   & 0.04  &  3   &    0.06   & 0.00  &  2&       0.06   & 5883  &  4.51  &  1.18\\
\object{HD\,150706} &-0.08   &0.03    &3     &  0.01   & 0.06  &  3   &    -0.12  & 0.03  &  3&       -0.01  & 5961  &  4.50  &  1.11\\
\object{HD\,160691} &0.38    &0.04    &3     &  0.35   & 0.04  &  3   &    0.39   & 0.02  &  2&       0.32   & 5798  &  4.31  &  1.19\\
\object{HD\,162020} &-0.28   &0.04    &2     &  0.06   & 0.04  &  3   &    0.05   & 0.00  &  2&       -0.04  & 4858  &  4.42  &  0.86\\
\object{HD\,168443} &0.04    &0.01    &2     &  0.17   & 0.03  &  3   &    0.24   & 0.02  &  3&       0.06   & 5617  &  4.22  &  1.21\\
\object{HD\,168746} &-0.04   &0.04    &3     &  0.18   & 0.02  &  3   &    0.23   & 0.04  &  2&       -0.08  & 5601  &  4.41  &  0.99\\
\object{HD\,169830} &0.41    &0.01    &2     &  0.18   & 0.05  &  2   &    0.18   & 0.02  &  2&       0.21   & 6299  &  4.10  &  1.42\\
\object{HD\,177830} &0.37    &0.03    &2     &  0.56   & 0.05  &  3   &    0.54   & 0.04  &  2&       0.33   & 4804  &  3.57  &  1.14\\
\object{HD\,178911} &0.12    &0.06    &2     &  0.30   & 0.04  &  3   &    0.27   & 0.00  &  1&       0.27   & 5600  &  4.44  &  0.95\\
\object{HD\,179949} &0.30    &0.01    &3     &  0.22   & 0.05  &  2   &    n.d.   & n.d.  &  0&       0.22   & 6260  &  4.43  &  1.41\\
\object{HD\,186427} &0.07    &0.00    &1     &  0.12   & 0.05  &  3   &    0.11   & 0.03  &  3&       0.08   & 5772  &  4.40  &  1.07\\
\object{HD\,187123} &0.15    &0.00    &1     &  0.14   & 0.01  &  3   &    0.13   & 0.01  &  3&       0.13   & 5845  &  4.42  &  1.10 \\
\object{HD\,190228} &-0.24   &0.01    &3     &  -0.16  & 0.02  &  3   &    -0.12  & 0.00  &  1&       -0.26  & 5327  &  3.90  &  1.11\\
\object{HD\,190360} &0.26    &0.04    &2     &  0.33   & 0.04  &  4   &    0.34   & 0.04  &  4&       0.24   & 5584  &  4.37  &  1.07\\
\object{HD\,192263} &n.d.    &n.d.    &0     &  0.07   & 0.05  &  3   &    0.02   & 0.01  &  2&       -0.02  & 4947  &  4.51  &  0.86\\
\object{HD\,195019} &0.04    &0.02    &3     &  0.13   & 0.02  &  3   &    0.19   & 0.01  &  2&       0.09   & 5842  &  4.32  &  1.27\\
\object{HD\,196050} &0.34    &0.02    &3     &  0.31   & 0.04  &  3   &    0.40   & 0.02  &  2&       0.22   & 5918  &  4.35  &  1.39\\
\object{HD\,202206} &0.38    &0.02    &2     &  0.41   & 0.06  &  2   &    0.37   & 0.00  &  2&       0.35   & 5752  &  4.50  &  1.01\\
\object{HD\,209458} &-0.04   &0.00    &2     &  0.01   & 0.02  &  2   &    0.00   & 0.01  &  2&       0.02   & 6117  &  4.48  &  1.40 \\
\object{HD\,210277} &0.2     &0.03    &3     &  0.32   & 0.05  &  3   &    0.35   & 0.02  &  2&       0.19   & 5532  &  4.29  &  1.04\\
\object{HD\,213240} &0.22    &0.04    &3     &  0.23   & 0.03  &  3   &    0.20   & 0.00  &  1&       0.17   & 5984  &  4.25  &  1.25\\
\object{HD\,216435} &0.37    &0.01    &3     &  0.31   & 0.04  &  3   &    0.32   & 0.01  &  2&   0.24   & 5938  &  4.12  &  1.28\\
\object{HD\,216437} &0.35    &0.05    &3 &  0.31   & 0.02  &  3   &    0.40   & 0.02  &  2&	  0.25   & 5887  &  4.30  &  1.31\\
\object{HD\,216770} &0.18    &0.00    &1 &  0.43   & 0.00  &  1   &    0.37   & 0.05  &  4&	  0.26   & 5423  &  4.40  &  1.01\\
\object{HD\,217014} &0.25    &0.04    &3 &  0.3    & 0.03  &  3   &    0.31   & 0.01  &  2&	  0.20   & 5804  &  4.42  &  1.20 \\
\object{HD\,217107} &0.38    &0.04    &3 &  0.41   & 0.05  &  3   &    0.44   & 0.02  &  2&	  0.37   & 5646  &  4.31  &  1.06\\
\object{HD\,222404} &0.12    &0.00    &1 &  0.34   & 0.04  &  2   &    0.40   & 0.04  &  3&	  0.16   & 4916  &  3.36  &  3.36\\
\object{HD\,222582} &0.05    &0.02    &3 &  0.16   & 0.02  &  3   &    0.15   & 0.03  &  2&	  0.05   & 5843  &  4.45  &  1.03\\
\object{HD\,330075} &-0.05   &0.02    &2 &  0.17   & 0.00  &  2   &    0.35   & 0.01  &  2&	  0.08   & 5017  &  4.22  &  0.69\\
\hline
\end{tabular}
\label{tab:planets2}										     
\end{table*}

\newpage

\begin{table*}[t!]
\center
\caption{Same as Table\,\ref{tab:planets1} for the comparison sample of stars without known planets.}
\begin{tabular}{lccccccccccccc}
\hline
Star & [Na/H] & $\sigma$ & $n$ & [Mg/H] & $\sigma$ & $n$ & [Al/H] & $\sigma$ & $n$ & [Fe/H] & T$_{eff}$ & log$g$ & $\xi_t$\\
\hline
\object{HD\,1581  }   &-0.10   &0.05	&3     &  -0.02  & 0.00  &  2  &     -0.14  & 0.01    &2      & -0.14 &  5956 &   4.39 &   1.07\\
\object{HD\,4391  }   &-0.09   &0.05	&3     &  0.10   & 0.01  &  2  &     -0.02  & 0.00    &1      & -0.03 &  5878 &   4.74 &   1.13\\
\object{HD\,5133  }   &-0.27   &0.04	&2     &  -0.03  & 0.04  &  3  &     -0.07  & 0.01    &2      & -0.17 &  4911 &   4.49 &   0.71\\
\object{HD\,7570  }   &0.24    &0.03	&3     &  0.22   & 0.01  &  3  &     0.20   & 0.01    &2      & 0.18  &  6140 &   4.39 &   1.5 \\
\object{HD\,10360 }   &-0.29   &0.05	&2     &  -0.13  & 0.04  &  3  &     -0.13  & 0.01    &2      & -0.26 &  4970 &   4.49 &   0.76\\
\object{HD\,10700 }   &-0.52   &0.03	&2     &  -0.23  & 0.01  &  3  &     -0.24  & 0.02    &2      & -0.52 &  5344 &   4.57 &   0.91\\
\object{HD\,14412 }   &-0.42   &0.04	&3     &  -0.3   & 0.03  &  3  &     -0.34  & 0.03    &2      & -0.47 &  5368 &   4.55 &   0.88\\
\object{HD\,17925 }   &-0.03   &0.05	&3     &  0.20   & 0.04  &  2  &     0.08   & 0.01    &2      & 0.06  &  5180 &   4.44 &   1.33\\
\object{HD\,20010 }   &-0.10   &0.03	&3     &  -0.13  & 0.05  &  2  &     n.d.   & n.d.    &0      & -0.19 &  6275 &   4.4  &   2.41\\
\object{HD\,20766 }   &-0.23   &0.01	&3     &  -0.02  & 0.01  &  2  &     -0.08  & 0.02    &2      & -0.21 &  5733 &   4.55 &   1.09\\
\object{HD\,20794 }   &-0.28   &0.04	&3     &  -0.05  & 0.03  &  3  &     -0.03  & 0.02    &2      & -0.38 &  5444 &   4.47 &   0.98\\
\object{HD\,20807 }   &-0.24   &0.02	&3     &  -0.11  & 0.05  &  3  &     -0.18  & 0.03    &2      & -0.23 &  5843 &   4.47 &   1.17\\
\object{HD\,23249 }   &0.12    &0.05	&3     &  0.22   & 0.04  &  3  &     0.25   & 0.03    &2      & 0.13  &  5074 &   3.77 &   1.08\\
\object{HD\,23356 }   &-0.26   &0.01	&0     &  0.00   & 0.04  &  3  &     0.00   & 0.01    &2      & -0.11 &  4975 &   4.48 &   0.77\\
\object{HD\,23484 }   &-0.06   &0.06	&3     &  0.12   & 0.04  &  3  &     0.09   & 0.02    &2      & 0.06  &  5176 &   4.41 &   1.03\\
\object{HD\,26965 }   &-0.26   &0.02	&2     &  -0.02  & 0.03  &  3  &     0.00   & 0.01    &2      & -0.31 &  5126 &   4.51 &   0.60\\
\object{HD\,30495 }   &-0.05   &0.03	&3     &  0.05   & 0.03  &  3  &     0.05   & 0.01    &2      & 0.02  &  5868 &   4.55 &   1.24\\
\object{HD\,36435 }   &-0.12   &0.05	&3     &  0.04   & 0.01  &  3  &     -0.03  & 0.02    &2      & 0.00  &  5479 &   4.61 &   1.12\\
\object{HD\,38858 }   &-0.22   &0.01	&3     &  -0.13  & 0.04  &  3  &     -0.13  & 0.00    &2      & -0.23 &  5752 &   4.53 &   1.26\\
\object{HD\,40307 }   &n.d.    &n.d.	&0     &  -0.15  & 0.05  &  3  &     -0.14  & 0.04    &2      & -0.3  &  4805 &   4.37 &   0.49\\
\object{HD\,43162 }   &-0.08   &0.04	&3     &  0.09   & 0.00  &  2  &     0.02   & 0.02    &2      & -0.01 &  5633 &   4.48 &   1.24\\
\object{HD\,43834 }   &0.09    &0.02	&3     &  0.25   & 0.02  &  3  &     0.24   & 0.05    &2      & 0.10  &  5594 &   4.41 &   1.05\\
\object{HD\,50281 }   &-0.36   &0.04	&2     &  0.04   & 0.06  &  3  &     0.00   & 0.01    &2      & -0.04 &  4658 &   4.32 &   0.64\\
\object{HD\,53705 }   &-0.16   &0.02	&3     &  -0.04  & 0.04  &  3  &     -0.04  & 0.01    &2      & -0.19 &  5825 &   4.37 &   1.20\\
\object{HD\,53706 }   &-0.26   &0.04	&3     &  -0.07  & 0.00  &  2  &     -0.01  & 0.00    &2      & -0.26 &  5260 &   4.35 &   0.74\\
\object{HD\,65907 }   &-0.19   &0.04	&3     &  0.03   & 0.03  &  2  &     0.00   & 0.00    &2      & -0.29 &  5979 &   4.59 &   1.36\\
\object{HD\,69830 }   &-0.09   &0.04	&3     &  0.08   & 0.00  &  2  &     0.11   & 0.01    &2      & -0.03 &  5410 &   4.38 &   0.89\\
\object{HD\,72673 }   &-0.37   &0.04	&3     &  -0.22  & 0.06  &  3  &     -0.22  & 0.02    &2      & -0.37 &  5242 &   4.50 &   0.69\\
\object{HD\,74576 }   &-0.15   &0.04	&2     &  0.04   & 0.04  &  3  &     -0.03  & 0.01    &2      & -0.03 &  5000 &   4.55 &   1.07\\
\object{HD\,76151 }   &0.13    &0.02	&3     &  0.20   & 0.05  &  3  &     0.16   & 0.03    &2      & 0.14  &  5803 &   4.50 &   1.02\\
\object{HD\,84117 }   &0.05    &0.05	&3     &  0.05   & 0.05  &  2  &     n.d.   & n.d.    &0      & -0.03 &  6167 &   4.35 &   1.42\\
\object{HD\,189567}   &-0.24   &0.02	&3     &  0.00   & 0.04  &  2  &     -0.07  & 0.00    &1      & -0.23 &  5765 &   4.52 &   1.22\\
\object{HD\,191408}   &-0.46   &0.06	&3     &  -0.24  & 0.00  &  2  &     -0.14  & 0.02    &2      & -0.55 &  5005 &   4.38 &   0.67\\
\object{HD\,192310}   &0.00    &0.03	&2     &  0.09   & 0.02  &  2  &     0.13   & 0.00    &2      & -0.01 &  5069 &   4.38 &   0.79\\
\object{HD\,196761}   &-0.3    &0.03	&3     &  -0.13  & 0.03  &  3  &     -0.13  & 0.02    &2      & -0.29 &  5435 &   4.48 &   0.91\\
\object{HD\,207129}   &-0.03   &0.03	&3     &  0.05   & 0.03  &  3  &     0.01   & 0.01    &2      & 0.00  &  5910 &   4.42 &   1.14\\
\object{HD\,209100}   &n.d.    &n.d.	&0     &  -0.05  & 0.01  &  2  &     -0.04  & 0.01    &2      & -0.06 &  4629 &   4.36 &   0.42\\
\object{HD\,211415}   &-0.13   &0.04	&3     &  -0.04  & 0.02  &  2  &     -0.09  & 0.04    &2      & -0.17 &  5890 &   4.51 &   1.12\\
\object{HD\,216803}   &n.d.    &n.d.	&0     &  0.01   & 0.05  &  3  &     -0.09  & 0.03    &2      & -0.01 &  4555 &   4.53 &   0.66\\
\object{HD\,222237}   &n.d.    &n.d.	&0     &  -0.07  & 0.02  &  2  &     -0.07  & 0.03    &2      & -0.31 &  4747 &   4.48 &   0.40\\
\object{HD\,222335}   &-0.22   &0.03    &3     &  -0.06  & 0.04  &  3  &     -0.11  & 0.01    &2      & -0.16 &  5260 &   4.45 &   0.92\\
\hline													     
\end{tabular}	
\label{tab:comparison}										     
\end{table*}

\end{document}